\documentclass[prc,amsmath,twocolumn,showpacs,superscriptaddress]{revtex4}
\usepackage[dvips]{graphicx}
\usepackage{dcolumn}

\newcommand{\bea}{\begin{eqnarray}}
\newcommand{\eea}{\end{eqnarray}}
\newcommand{\be}{\begin{equation}}
\newcommand{\ee}{\end{equation}}

\newcommand{\fig}[1]{Fig.~\ref{#1}}
\newcommand{\figs}[2]{Figs.~\ref{#1} and \ref{#2}}
\renewcommand{\vec}[1]{\boldsymbol{#1}}
\newcommand{\bra}[1]{\langle#1|}
\newcommand{\ket}[1]{|#1\rangle}

\newcommand{\Rr}{\vec{R},\vec{r}}

\begin{document}

\title{Sensitivity of $^8$B breakup cross section to projectile structure in CDCC calculations}

\author{N.~C.~Summers}
 \email{summers@nscl.msu.edu}
 \affiliation{National Superconducting Cyclotron Laboratory, 
Michigan State University, East Lansing, Michigan 48824}
\author{F.~M.~Nunes}
 \affiliation{National Superconducting Cyclotron Laboratory, 
Michigan State University, East Lansing, Michigan 48824}
 \affiliation{Department of Physics and Astronomy, 
Michigan State University, East Lansing, Michigan 48824}
\date{\today}

\begin{abstract}
Given the Astrophysical interest of $^7$Be$(p,\gamma)^8$B, there have been
several experiments applying the Coulomb dissociation method for extracting
the capture rate.
Measurements at Michigan State are dominated by $E1$ contributions
but have  a small $E2$ component.
On the other hand, a lower energy measurement at Notre Dame 
has a much stronger $E2$ contribution. The expectation was
that the two measurements would tie down the $E2$ and thus allow for
an accurate extraction of the $E1$ relevant for the capture process.
The aim of this brief report is to show that the $E2$ factor in breakup
reactions does not translate into a scaling of the $E2$ contribution in
the corresponding capture reaction.
We show that changes to the $^8$B single particle parameters, which
are directly related to the $E2$ component in the capture reaction, do not
effect the corresponding breakup reactions, using the present reaction theory.
\end{abstract}

\pacs{24.10.Eq, 25.60.Gc, 25.70.De, 27.20.+n}

\maketitle

Breakup reactions are one of the best probes to study 
nuclei on the dripline. The loosely bound nature of the nuclear systems imply
that the continuum plays a very important role in the reaction mechanism.
In the past few years, the Continuum Discretized Coupled Channel method
\cite{cdcc-theory} has been successfully applied to exotic nuclei. This method
is fully quantum mechanical and is non-perturbative: it includes the couplings
to breakup states to all orders. The projectile is treated as a $core+N$ system
which means that one of the main inputs to the model is the Hamiltonian of the 
projectile generating both bound and scattering states. In order
to be able to treat continuum-continuum couplings, the scattering states are
bunched into energy bins. Apart from the $core+N$ interaction, the CDCC method
also requires the optical potentials for $core+target$ and $N+target$ which are
typically well known. 

One of the first applications of CDCC to dripline nuclei involved the description of
the breakup of $^8$B on $^{58}$Ni at $E_{\mathrm{beam}}=26.5$ MeV \cite{nunes-99}. 
The corresponding experiment was performed at Notre Dame \cite{kolata-01}. 
The calculations in \cite{nunes-99} involved no fitting. 
The potential model for $^8$B$=^7$Be$+p$  was taken from \cite{esbensen-96}.
It assumes that, in the ground state, 
the valence proton in $^8$B is a single particle $1p_{3/2}$
coupled to the ground state of the $^7$Be core. 
Note that a simplified version of the interaction from \cite{esbensen-96}
was used in order to have the same interaction in all partial waves
in the continuum as the interaction for the ground state. 
In \cite{tostevin-01}, three body observables are calculated and integrated to enable a
direct comparison to the experimental data. 
The optical potentials for $p$-$^{58}$Ni and $^7$Be-$^{58}$Ni are the same as those
in \cite{nunes-99,tostevin-01}. 
The agreement for the angular and energy 
distributions of the heavy fragment was extremely good.

The experiment \cite{kolata-01} measured only the heavy fragment $^7$Be
which means that the data contains both breakup (which is also  referred to as
diffraction) and stripping. However it was shown \cite{esbensen-99} that 
stripping only contributes for larger angles. The CDCC calculations
performed make predictions for breakup only. Therefore our discussion will 
focus on the smaller angles only.

Another experiment at MSU measured the breakup of $^8$B on Au and Pb at 40-80 MeV/A 
\cite{davids-98,davids-01}.
% Both fragments ($p$ and $^7$Be) were detected.
Amongst other observables, detailed momentum distributions for $^7$Be were extracted.
CDCC calculations for these data were performed in \cite{mortimer-02}. 
Starting with the same
$^7$Be+$p$ interaction as in \cite{nunes-99}, results show
that good agreement with the data for the various angular sets and both
targets can only be obtained if the quadrupole excitation
couplings are artificially increased by a factor of 1.6.

In this brief report we wish to highlight the differences between the
quadrupole strength of higher-order reaction theory and the $E2$ capture
strength of the astrophysically relevant inverse reaction.
We will examine how the $1.6$ factor from Ref.~\cite{mortimer-02}
translates into the $E2$ component of $^7$Be$(p,\gamma)^8$B.
For this purpose, we have repeated the calculations for both
breakup reactions with a modified $^7$Be+$p$ interaction.
We show that this interaction, chosen to have a drastic effect on the
$E2$ capture cross section, has a minimal effect on both breakup reactions.
In addition, we discuss the implications of the findings 
in \cite{mortimer-02} for the Notre Dame experiment,
and show that the results of both experiments cannot be consistently described
within the same CDCC single particle model used until now.

It is important to clarify the difference between
the quadrupole strength used in breakup reactions and the
$E2$ strength of the astrophysically relevant capture.
In perturbation theory, the link is direct,
but in a fully quantum mechanical description of the scattering,
this relationship is not so transparent.
In CDCC, the excitation operator for the breakup reaction is the
coupling potential for the sum of the core-target
and proton-target interactions, expanded in multipoles,
\be
V(\Rr) = \sum_K (2K+1) V^K(R,r) P_K(\vec{r}\cdot\vec{R}),
\ee
where $P_K$ are the Legendre polynomials.
These multipoles are then averaged over each bin state, 
generated by a single particle model for $p$-$^7$Be,
\be
V^K_{\alpha:\alpha'}(R) = \bra{\phi_{\alpha'}(\vec{r})}
V^K(R,r) \ket{\phi_{\alpha}(\vec{r})}
\ee
Different optical potentials for the fragments and the target, 
as well as different $p$-$^7$Be wave functions, modify these
couplings.
The quadrupole strength in CDCC is the strength of these couplings,
which include both nuclear and Coulomb contributions,
and can have multistep effects.
Any modification to the $^8$B single particle parameters will effect
the breakup cross section through these couplings.
The inclusion of multistep and nuclear effects complicates the link between
the magnitude of the couplings and the cross section.
In contrast, the $E2$ operator in capture reactions comes from
the expansion of the electro-magnetic field and is very well understood 
\cite{esbensen-96}.
The cross section is obtained directly from the $E2$ operator.
Differences in $E2$ components for the capture reaction translate
directly into differences in the $p$-$^7$Be interaction \cite{nunes-98}.

\begin{figure}
\includegraphics[width=8cm]{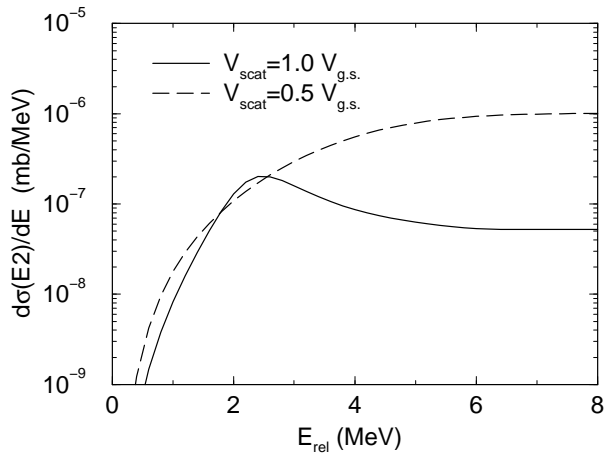}
\caption{\label{fig:cap} Capture cross section for  $^7$Be$(p,\gamma)^8$B:
sensitivity to the initial $^7$Be-$p$ interaction. }
\end{figure}
First we consider the effect of the $^8$B single particle parameters.
Given that in \cite{nunes-99} simplifications were made to the 
$p$-$^7$Be interaction in the continuum, we have explored the dependence 
of the breakup observables on this final state interaction.
A slight modification of the strength of the p-wave is needed
to reproduce the correct $1^+$ resonance. Even though this resonance
dominates the direct capture cross section, it has been shown to
produce no effect on the Coulomb Dissociation cross sections \cite{nunes-98}.
Several modifications to the interaction generating 
the scattering waves that can be reached through $E2$ 
(p- and f-waves) were explored (using the methods of Ref.~\cite{b8exc}).
The strongest modification of the $E2$ capture 
cross section was obtained when p- and f-waves are generated by a 
potential half as deep as the ground state potential (\fig{fig:cap}). 
This is completely unphysical for the breakup reaction in two ways:
the resonance structure that is known is not reproduced and the ground
state becomes non-orthogonal to the p-wave continuum states.
We only perform these calculations, in this extreme case,
to highlight the differences between the quadrupole couplings
in breakup and the $E2$ capture cross section. 
In \fig{fig:cap}, we show the results for the $E2$ component of 
capture cross section for $^7$Be$(p,\gamma)^8$B: the solid line refers
to the structure model where the depth is the same for bound and
continuum states (as in \cite{nunes-99}) and the dashed line where 
the potential depth in p- and f-wave scattering states is reduced to 
one half of its original value.

\begin{figure}[t]
\includegraphics[width=8cm]{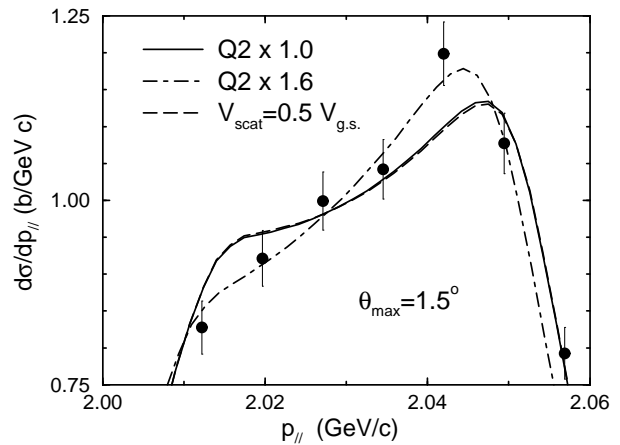}
\caption{\label{fig:msu1}  Momentum distribution of $^7$Be following 
the breakup of 44 MeV/A $^8$B on a Pb target integrated
up to $\theta_{\mathrm{max}}=1.5^{\circ}$:
the previous results \cite{mortimer-02} with no renormalization
of the quadrupole excitation (solid line) and with a 1.6 
renormalization (dot-dashed line) and the 
result using 0.5*V($p$-$^7$Be) for the final state (dashed line). }
\end{figure}
\begin{figure}[t]
\includegraphics[width=8cm]{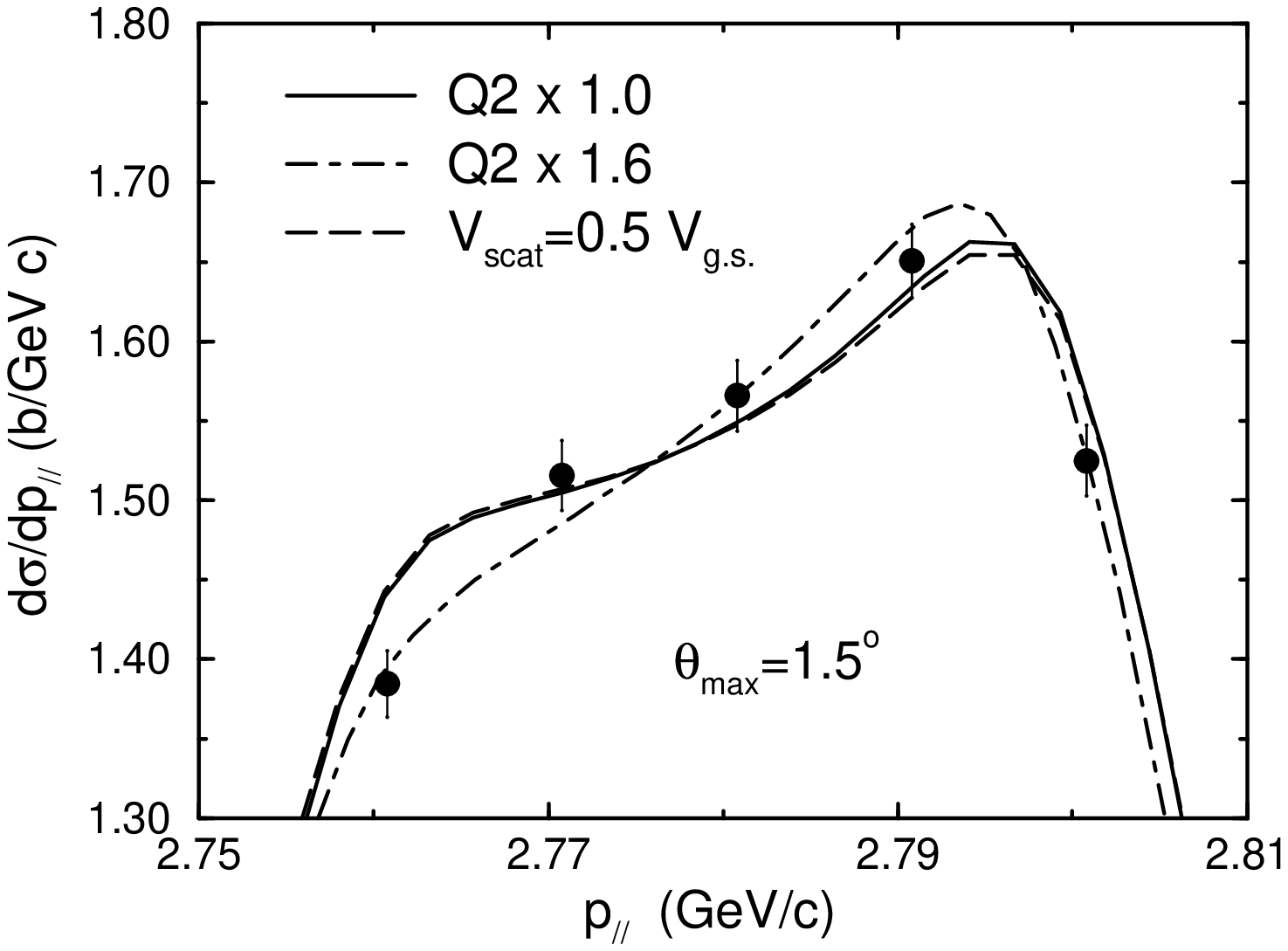}
\caption{\label{fig:msu2}  Momentum distribution of $^7$Be following 
the breakup of 81 MeV/A $^8$B on a Pb target integrated
up to $\theta_{\mathrm{max}}=1.5^{\circ}$. The lines correspond
to the same as \fig{fig:msu1}.}
\end{figure}
\begin{figure}
\includegraphics[width=8cm]{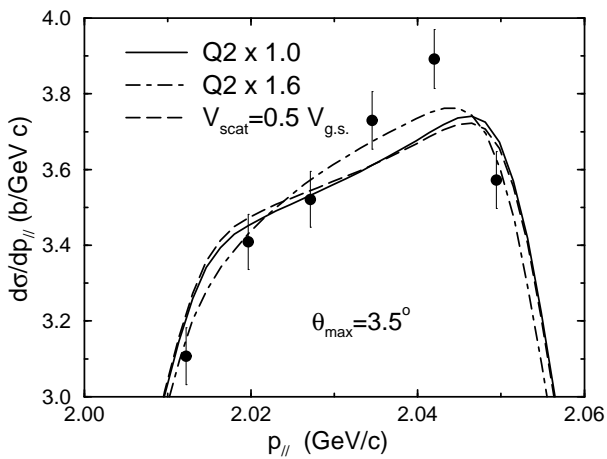}
\caption{\label{fig:msu3}  Momentum distribution of $^7$Be following 
the breakup of 44 MeV/A $^8$B on a Pb target 
integrated up to $\theta_{\mathrm{max}}=3.5^{\circ}$.
The lines correspond to the same as \fig{fig:msu1}.}
\end{figure}
We now apply this modified $^8$B interaction to the MSU breakup data.
We reproduce the calculations of Ref.~\cite{mortimer-02},
using the code {\sc fresco} \cite{fresco},
within the same model space for the CDCC calculations, keeping
the same optical potentials, and the same single particle parameters
for the s- and d-waves, but with our modified interaction for the bound state,
p- and f-waves in the continuum.
We chose the Pb target as an example and performed the
calculations at both 44 MeV/A (\fig{fig:msu1}) and 81 MeV/A (\fig{fig:msu2}).
The solid line corresponds to previous calculations
with no artificial factors on the quadrupole strength;
the dot-dashed line corresponds to the results of solving
the CDCC equations with a quadrupole excitation multiplied by 1.6
and the dashed line is the result when the interaction for
the p- and f-waves of the $p$-$^7$Be system is reduced to
one half without the additional 1.6 factor in the quadrupole couplings. 

The higher the beam energy, the more sensitive one is to the higher 
$p$-$^7$Be excitation energies and, as the $E2$ capture is mostly different
for the higher excitation energies, one might expect the 44 MeV/A and 81 MeV/A
reaction to be sensitive to this change even though the $E2$/$E1$ cross section ratio
decreases with beam energy. However, data was only
taken at forward angles, which effectively imposes small relative
energies.
With the angular truncation of $\theta_{\mathrm{max}}=1.5^{\circ}$ there
is virtually no effect due to the change of the $p$-$^7$Be scattering potential
(\figs{fig:msu1}{fig:msu2})
and a small effect can be seen when angles up to $\theta_{\mathrm{max}}=3.5^{\circ}$
are considered (\fig{fig:msu3}). 

Albeit the large modification on the $E2$ capture cross sections,
the modification of the interaction has an insignificant effect
on the breakup reaction observable here.
This clearly shows the difference between quadrupole strength in
higher order reaction theory and the $E2$ capture strength.

\begin{figure}[t]
\includegraphics[width=8cm]{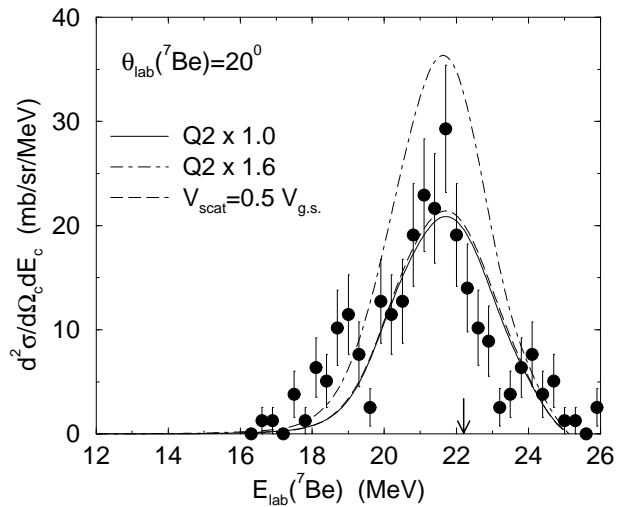}
\caption{\label{FIG:en} Energy distribution of $^7$Be at 20 degrees. 
Details of the presented curves can be found in the text. }
\end{figure}
\begin{figure}[b]
\includegraphics[width=8cm]{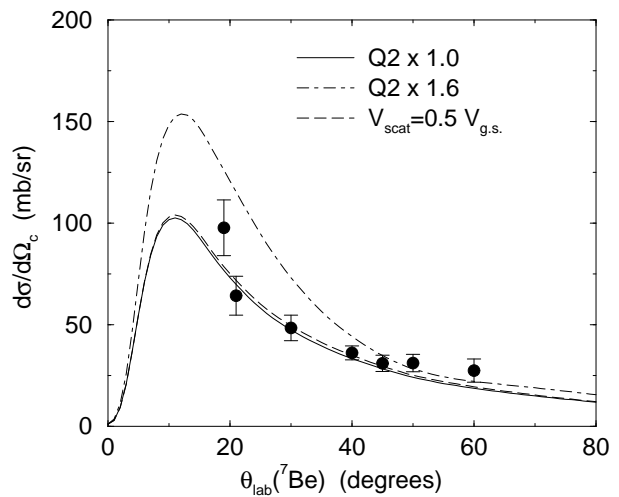}
\caption{\label{FIG:ang} Angular distribution of $^7$Be following 
the breakup of 25.6 MeV $^8$B on a $^{58}$Ni target. 
Details of the presented curves can be found in the text. }
\end{figure}

Now let us consider the Notre Dame experiment.
The energy distribution at $\theta_{\mathrm{lab}}$($^7$Be)=20$^{\circ}$ 
is displayed in \fig{FIG:en}, and the angular distribution is presented in
\fig{FIG:ang}.
We reproduce the calculations from \cite{tostevin-01},
but with our modified $^8$B single particle model for p- and f-waves (dashed line).
The data is taken from \cite{kolata-01}.
In both figures the solid line represents the calculations of Ref.~\cite{tostevin-01},
where no scaling of the quadrupole strength was introduced.
As with the MSU data, the modified $^8$B single particle potential,
which produced a large modification to the $E2$ capture strength,
has a negligible effect on the breakup cross section within the CDCC reaction model.

Given that quadrupole excitations have a larger
contribution in the lower energy reactions, the obvious question arising from
the work \cite{mortimer-02} is whether the error bars in the Notre Dame 
data could allow for this rather large increased quadrupole strength. Note that
in this low energy regime, Coulomb-nuclear and $E1$-$E2$ interference are very
strong and it is not clear how a larger quadrupole excitation would affect
the predictions.

We reproduce the calculations from \cite{tostevin-01} 
within the same model space, 
keeping the same optical potentials, 
and the single particle parameters for $^8$B \cite{esbensen-96} but 
multiplying the generated quadrupole strength by the factor of 1.6
as suggested by \cite{mortimer-02}. 
We compare it with the previous results, 
where no artificial increase is imposed. 
The coupled channel code {\sc fresco} \cite{fresco} was used.
The resulting energy distributions are presented in
\fig{FIG:ang} and angular distributions in \fig{FIG:en}.
It is clear from the figures that, by imposing such a large quadrupole term, one
no longer can describe the data satisfactorily.

The results at $20^\circ$ are not completely nuclear free and are certainly
influenced by multi-step effects \cite{nunes-99}.
Therefore the scaling of $1.6$ cannot be directly translated into a scaling of the cross section.
We have performed $E2$ only calculations for both the original $^8$B model and the scaled $1.6$ model.
The ratio of the $E2$ cross sections at the peak of the energy distribution for $20^{\circ}$ is 2.7.
The final numbers when $E1$ and nuclear are included show a $1.7$ ratio 
between the new calculations and the old.
This demonstrates the importance of interference and makes a reliable extraction of the $E2$ strength from a particular set of data less transparent although still possible.

Compared to this strong quadrupole effect, there is a much weaker dependence on the 
optical potentials.
In fact, the results in \cite{tostevin-01} 
are very weakly dependent on the $^7$Be-$^{58}$Ni interaction and
depend only slightly on the proton optical potential which is well known
(solid and dotted-dashed lines in Fig.~4a of \cite{tostevin-01}). 
Consequently, the disagreement would not disappear by readjusting optical potentials.
As to possible experimental problems, even if the Notre Dame data suffered from a $50$ \%
error in the absolute normalization, which is extremely improbable,
the energy distributions would be much broader than the model's prediction.

Given the series of exploratory calculations performed, we claim that
within the present single particle description of $^8$B, 
a $1.6$ factor in the quadrupole breakup couplings cannot be 
accounted by modifying the $E2$ component
of the $^7$Be$(p,\gamma)^8$B within a single particle $p$-$^7$Be picture. 
 
Altogether, these results suggest that there is additional physics 
not included in the present calculations relevant for the
breakup of this nucleus. For example, our breakup model uses a 
detailed description of the reaction process based on a simplistic
description of the structure of the projectile. Although it is standard practice
to assume that $^8$B is a single particle proton $p_{3/2}$ built on the ground state of $^7$Be, 
it is known that it contains a 15\% core excited component \cite{gsi}. It is not
understood how this component would dynamically interfere throughout the
reaction process. 
One possibility is that this component affects the MSU
data differently, enhancing the quadrupole excitation. At present, 
dynamical core excitation is not possible within a fully coupled quantum 
calculation. Future work on an extension of the standard CDCC method is needed.

\medskip
We thank Sam Austin, Jim Kolata, Jeff Tostevin and Ian Thompson 
for valuable comments on an earlier version of the manuscript.
We are indebted to Jeff Tostevin for providing the codes to
generate the momentum distributions.
This work is supported by NSCL, Michigan State University.

\end{document}